\documentclass[12pt]{iopart}

\begin{document}

\title[Eddington's Gravity in Immersed Spacetime]{Eddington's Gravity in Immersed Spacetime}

\author{Hemza Azri}

\address{Department of Physics, {\.I}zmir Institute of Technology, TR35430 {\.I}zmir, Turkey.}
\ead{hemzaazri@iyte.edu.tr}
\vspace{10pt}
\begin{indented}
\item[]January 2014
\end{indented}

\begin{abstract}
We formulate Eddington's affine gravity in a spacetime which is immersed in a larger eight dimensional space endowed with a hypercomplex structure. The dynamical equation of the first immersed Ricci-type tensor leads to gravitational field equations which include matter. We also study the dynamical effects of the second Ricci-type tensor when added to the Lagrangian density. A simple Lagrangian density constructed from combination of the standard Ricci tensor and a new tensor field that appears due to the immersion, leads to gravitational equations in which the vacuum energy gravitates with a different cosmological strength as in Phys. Rev. D {\bf 90}, 064017 (2014), rather than with Newton's constant. As a result, the tiny observed curvature is reproduced due to large hierarchies rather than fine-tuning.    
\end{abstract}

%
%
%
%
%

\section{Introduction}

General Relativity is the relativistic theory of gravity where its field equations are derived from the variation of Einstein-Hilbert action with respect to the fundamental field, \textit{the metric tensor} \cite{EH1,EH2,EH3,EH4}. This theory is based on a purely metric formulation where the affine connection envisaged in the space is the Levi-Civita connection of that metric. 
\newline
However, the purely metric formulation is not the only way to construct a theory of gravity. In fact, it has been noticed that rather than metric, General Relativity is based on the affine connection \cite{Weyl}. The simplest formulation of gravity based on affine connection is Eddington's gravity \cite{Ed1,Ed2}. In this theory, the field equations equivalent to Einstein's equations with only a cosmological constant are derived from a least action principle where the fundamental quantity is the affine connection \cite{Schrodinger}     
   
In Eddington's purely affine gravity where the connection is taken symmetric with no notion of metric, the gravitational field equations are derived from the principle of least action where the covariant Lagrangian density is constructed by the square root of the symmetric part of the covariant Ricci tensor as follows
\begin{equation}
S_{Edd}=\int d^{4}x \sqrt{\texttt{Det}\left[ \mathcal{R}\right]},
\end{equation}
where $\texttt{Det}\left[ \mathcal{R}\right]$ is the determinant of the symmetric part of the Ricci tensor $\mathcal{R_{\alpha\beta}}$. We note that in this work, the non-symmetric part of the Ricci tensor is assumed to be zero.
\newline
The variation of the Ricci tensor with respect to the connection $\Gamma$ is
\begin{equation}
\delta {\mathcal{R}}_{\alpha\beta} = \nabla_{\mu}\Big( \delta\Gamma^{\mu}_{\beta\alpha}\Big) -
\nabla_{\beta}\Big(\delta\Gamma^{\mu}_{\mu\alpha}\Big),
\end{equation}
and then the principle of least action $\delta S_{Edd}=0$ leads to the equation
\begin{equation}
\nabla_{\mu} \left[ \sqrt{{\texttt{Det}}\left[{\mathcal{R}}\right]}
\left({\mathcal{R}}^{-1}\right)^{\alpha\beta}\right]=0. \label{motion0}
\end{equation}
This equation is solved by introducing an invertible and covariantly-constant tensor field $g_{\alpha\beta}$ such that
\begin{equation}
\sqrt{{\texttt{Det}}\left[{\mathcal{R}}\right]}
\left({\mathcal{R}}^{-1}\right)^{\alpha\beta}  = \lambda
\sqrt{g} g^{\alpha\beta},
\end{equation}
where $\lambda$ is a constant and $g = {\texttt{Det}}\left[g_{\alpha\beta}\right]$.
\newline
The last equation can be rewritten as Einstein's field equations with a cosmological constant $\lambda$
\begin{equation}
{\mathcal{R}}_{\alpha\beta} = \lambda g_{\alpha\beta}. \label{Edd}
\end{equation}
The compatibility condition $\nabla_{\mu} g_{\alpha\beta} =0$ which has its origin now from the equation of motion (\ref{motion0}) defines completely the Levi-Civita connection 
\begin{equation}
{}^{g}\Gamma^{\mu}_{\alpha\beta}= \frac{1}{2} g^{\mu\lambda} \left( \partial_{\alpha} g_{\beta\lambda} +
\partial_{\beta} g_{\lambda\alpha} - \partial_{\lambda} g_{\alpha\beta}\right).
\end{equation}

As we see, based on an affine connection Eddington's approach reproduces clearly the field equations of General Relativity in vacuum (\ref{Edd}) with its metric structure. In spite of being of great interest both physically and mathematically, Eddington's affine theory of gravity is considered incomplete as it does not include matter.

An ``Eddington-inspired Born-Infeld Gravity'' was proposed as an extension of Eddington's theory including matter \cite{Banados}. In this metric-affine formulation, the field equations are derived from a Lagrangian density where the variation is with respect to both quantities metric and affine connection, which are considered independent.   
\newline
Recently, it was shown that matter can be incorporated when Eddington's purely affine gravity is extended with Riemann curvature \cite{Durmus1}. In addition to incorporating matter, this ``Riemann-improved Eddington theory'' has enabled degravitation of the vacuum energy without any fine-tuning. The reason is that the cosmological constant was found to gravitate with a different \textit{cosmological strength} $M_{Co}$ rather than by Newton's constant.       

In this work we tackle the problem of incorporating matter in purely affine formulation. Our approach is based on spacetime which is considered to be plunged into a larger eight dimensional space which has a hypercomplex structure \cite{Clerc,Crum,AB}. As a result, in addition to an affine connection, the space became endowed with a new tensor of rank (2,1). We will show that the matter can be produced in the field equations as long as the Eddington's-like action is constructed from the Ricci-type tensors of this model. 
\newline
We will also show that a possible action constructed from combination of the standard Ricci tensor of the symmetric affine connection and the new tensor stated above, leads to degravitation of the cosmological constant as in \cite{Durmus1}. 

This paper is organized as follows: In Section 2, we review the mathematical construction which is used in this model. In Section 3, we construct the actions from the Ricci-type tensors and derive the field equations with matter. The idea of degravitation of the cosmological constant from this setup is given in Section 4. In Section 5 we give our conclusion.     

\section{Immersed Spacetime Structure}

The four dimensional spacetime $V_{4}$ is assimilated to an affine space with an affine connection, plunged into an
eight dimensional manifold $V_{8}$ which is the product of two
identical four dimensional real manifolds $W_{4}$ \cite{Clerc}
\begin{equation}
V_{8}=W_{4}\times W_{4}. \label{stru}
\end{equation}
We use the following convention \cite{Lichnerowicz,Yano}: 
\newline
For Latin indices: $i,j,..=1,...8$ and for Greek indices: $\alpha,\beta,..=1,...4$. We also introduce on the indices the operation $\ast$ such that $i^{\ast}=i\pm4$, (then $(i^{\ast})^{\ast}=i$).
\newline
This means that Latin indices take both Greek indices, $\alpha$ and $\alpha^{\ast}$ via the operation $\ast$, i.e, $i= \alpha, \alpha^{\ast}=1,...8$. 
\newline
It has been shown that this construction confers to the space $V_{8}$ a hypercomplex (or pseudo-complex) structure \cite{Clerc,Crum,AB}.
\newline
We define the hypercomplex coordinates $X^{\alpha}=x^{\alpha}+Ix^{\alpha^{\ast}}$ as elements of the Hypercomplex Ring $\mathbf{H}$, where $I^{2}=1$ and $x^{\alpha},x^{\alpha^{\ast}}$ are real coordinates from $W_{4}\times W_{4}$.
\newline
The diagonal submanifold $V_{4}$ is equivalent to
\cite{Clerc,AB}
\begin{equation}
x^{\alpha^{\ast}}=0.
\end{equation}
The real coordinates $x^{\alpha},x^{\alpha^{\ast}}$ are
called the associated diagonal coordinates.
\newline
As in the theory of complex manifolds, we define the almost hypercomplex structure on the tangent space of $V_{8}$ by the operator $J$ such that \cite{Lichnerowicz}
\begin{equation}
J\left(  \frac{\partial}{\partial x^{\alpha}}\right) = \frac{\partial}{\partial x^{\alpha^{\ast}}}, \quad J\left(  \frac{\partial}{\partial x^{\alpha^{\ast}}}\right) = \frac{\partial}{\partial x^{\alpha}}. 
\end{equation} 
This operator verifies $J^{2}=id$, with $id$ is the identity operator on the tangent space of $V_{8}$. 
\newline
In the real basis of $V_{8}$, this operator is defined by a tensor $J^{i}_{j}$ with its components given by the matrix \cite{Crum,Lichnerowicz}
\begin{equation}
\label{realmat}
J^{i}_{j}=
\left( {\begin{array}{cc}
0 & \mathcal{I}_{4} \\
\mathcal{I}_{4} & 0
\end{array} } \right),
\end{equation}
where $\mathcal{I}_{4}$ is the $4\times4$ unit matrix.
This means that $J$ has the components
\begin{equation}
\label{comp}
J^{\alpha}_{\beta} = J^{\alpha^{\ast}}_{\beta^{\ast}} =0, \quad
J^{\alpha}_{\beta^{\ast}} = J^{\alpha^{\ast}}_{\beta}= \delta^{\alpha}_{\beta}. 
\end{equation}
The operator $J$ corresponds to the multiplication by $I$ (with $I^{2}=1$). In fact, one may define a hypercomplex basis by the hypercomplex vectors
\begin{equation}
\frac{\partial}{\partial X^{\alpha}} = \frac{1}{2}\left(\frac{\partial}{\partial x^{\alpha}}+I\frac{\partial}{\partial x^{\alpha^{\ast}}}  \right), \quad \frac{\partial}{\partial X^{\alpha^{\ast}}} = \frac{1}{2}\left(\frac{\partial}{\partial x^{\alpha}}-I\frac{\partial}{\partial x^{\alpha^{\ast}}}  \right), 
\end{equation}
such that
\begin{equation}
J\left(  \frac{\partial}{\partial X^{\alpha}}\right) = I\frac{\partial}{\partial X^{\alpha}}, \quad J\left(  \frac{\partial}{\partial X^{\alpha^{\ast}}}\right) = -I\frac{\partial}{\partial X^{\alpha^{\ast}}}. 
\end{equation} 
The operator $J$ can be represented in the hypercomplex basis by the matrix
\begin{equation}
J^{i}_{j}=
\left( {\begin{array}{cc}
I\mathcal{I}_{4} & 0 \\
0 & -I\mathcal{I}_{4}
\end{array} } \right),
\end{equation}  
and the real representation of the linear group $GL(4,\mathbf{H})$ can be characterized as the subgroup of $GL(8,\mathbf{R})$ defined by the matrices which commute with (\ref{realmat}).  
\newline
In any frame of $V_{8}$, the connection form $\omega^{i}_{j}$ is represented by the matrix \cite{Lichnerowicz}
\begin{equation}
\omega^{i}_{j}=\left( {\begin{array}{cc}
\omega^{\alpha}_{\beta} & \omega^{\alpha}_{\beta^{\ast}} \\
\omega^{\alpha^{\ast}}_{\beta} & \omega^{\alpha^{\ast}}_{\beta^{\ast}}
\end{array} } \right).
\end{equation}   
The affine connection in the natural diagonal frame of $V_{8}$ is such that \cite{Clerc, Crum}
\begin{equation}
\omega^{\alpha}_{\beta} = \omega^{\alpha^{\ast}}_{\beta^{\ast}}, \quad \omega^{\alpha}_{\beta^{\ast}} = \omega^{\alpha^{\ast}}_{\beta}.
\end{equation}
In terms of components $\Gamma^{i}_{jk}$, the form $\omega^{i}_{j}$ is written locally as
\begin{equation}
\omega^{i}_{j} = \Gamma^{i}_{jk} dx^{k}.
\end{equation}
Now the affine connections in the natural diagonal frame bundle of $V_{8}$ are defined by the intrinsic conditions 
\begin{equation}
\label{condt}
\Gamma^{i}_{jk}=\Gamma_{j^{\ast}k}^{i^{\ast}},\quad \Gamma
_{jk}^{i}=\Gamma_{j^{\ast}k^{\ast}}^{i}.
\end{equation}
We can also derive the conditions (\ref{condt}) from the relation $\nabla J=0$, where $\nabla$ is the covariant derivative with respect to the connection $\Gamma_{jk}^{i}$, and $J$ is the operator of the almost hypercomplex structure given above by its components (\ref{comp}).
\newline 
When making the restriction in $V_{4}$, these conditions induce for all diagonal frame of $V_{4}$ the equations
\begin{equation}
\Gamma_{\beta\gamma}^{\alpha}=\Gamma_{\beta^{\ast}\gamma}^{\alpha^{\ast}
}=\Gamma_{\beta\gamma^{\ast}}^{\alpha^{\ast}}=\Gamma_{\beta^{\ast}\gamma
^{\ast}}^{\alpha},\quad \Gamma_{\beta\gamma}^{\alpha^{\ast}}
=\Gamma_{\beta^{\ast}\gamma}^{\alpha}=\Gamma_{\beta\gamma^{\ast}}^{\alpha
}=\Gamma_{\beta^{\ast}\gamma^{\ast}}^{\alpha^{\ast}}.\label{D7}
\end{equation}
We can show that the coefficients $\Gamma_{jk}^{i}$ with even number of
asterisks transform as connections, while those with odd number of
asterisks transform as tensors in all natural diagonal frame of
$V_{4}$ \cite{Lichnerowicz,Yano}.
Then according to Equations $(\ref{D7})$, one can define in $V_{4}$ an affine connection $\mathcal{L}_{\gamma\beta}^{\alpha}$ and a tensor $\Lambda
_{\beta\gamma}^{\alpha}$ by the following relations
\numparts
\begin{eqnarray}
\Gamma_{\beta\gamma}^{\alpha}=\Gamma_{\beta^{\ast}\gamma}^{\alpha^{\ast}
}=\Gamma_{\beta\gamma^{\ast}}^{\alpha^{\ast}}=\Gamma_{\beta^{\ast}\gamma
^{\ast}}^{\alpha}=\mathcal{L}_{\gamma\beta}^{\alpha},\\
\Gamma_{\beta\gamma}^{\alpha^{\ast}}=\Gamma_{\beta^{\ast}\gamma}^{\alpha
}=\Gamma_{\beta\gamma^{\ast}}^{\alpha}=\Gamma_{\beta^{\ast}\gamma^{\ast}
}^{\alpha^{\ast}}=\Lambda_{\beta\gamma}^{\alpha}, \label{D8}
\end{eqnarray}
\endnumparts
where the affine connection $\mathcal{L}_{\beta\gamma}^{\alpha}$ is generally asymmetric.
\newline
By the immersion of the submanifold $V_{4}$ in the manifold
$V_{8}$, the curvature form induced in $V_{4}$ is
\begin{equation}
\widehat{\Omega}_{j}^{i}
=\frac{1}{2}\widehat{R}_{j\lambda\mu}^{i}dx^{\lambda} \wedge dx^{\mu},
\end{equation}
where $\widehat{}$ \ means the restriction in $V_{4}$ (remember
that $x^{\mu^{\ast}}=0$ in $V_{4}$) and $R_{j\lambda\mu}^{i}$ is the Riemann tensor.
Then the induced Riemann tensor in $V_{4}$ becomes
\begin{equation}
\widehat{R}_{j\lambda\mu}^{i}=\partial_{\lambda}\Gamma_{j\mu}
^{i}-\partial_{\mu}\Gamma_{j\lambda}^{i}+\Gamma_{\rho\lambda}^{i}
\Gamma_{j\mu}^{\rho}+\Gamma_{j\mu}^{\rho^{\ast}}\Gamma
_{\rho^{\ast}\lambda}^{i}-\Gamma_{\rho\mu}^{i}\Gamma_{j\lambda}^{\rho}
-\Gamma_{\rho^{\ast}\mu}^{i}\Gamma_{j\lambda}^{\rho^{\ast}}.
\label{D19}
\end{equation}
From the induced Riemann tensor (\ref{D19}), one may construct only two independent Ricci-type tensors by contraction as follows
\begin{equation}
\mathcal{P}_{\alpha\beta}=\widehat{R}_{\beta\lambda\alpha}^{\lambda}, \quad
\mathcal{Q}_{\alpha\beta}=\widehat{R}_{\alpha^{\ast}\lambda\beta}^{\lambda}.
\end{equation} 
These tensors are calculated explicitly as \cite{Clerc,AB}
\numparts
\begin{eqnarray}
\mathcal{P}_{\alpha\beta}=\partial
_{\lambda}\mathcal{L}_{\alpha\beta}^{\lambda}-\partial_{\alpha}\mathcal{L}
_{\lambda\beta}^{\lambda}+\mathcal{L}_{\lambda\rho}^{\lambda}\mathcal{L}
_{\alpha\beta}^{\rho}-\mathcal{L}_{\alpha\rho}^{\lambda}\mathcal{L}
_{\lambda\beta}^{\rho}+\Lambda_{\rho\lambda}^{\lambda}\Lambda_{\beta\alpha}^{\rho}
-\Lambda_{\rho\alpha}^{\lambda}\Lambda_{\beta\lambda}^{\rho},\label{P}\\
\mathcal{Q}_{\alpha\beta}
=\partial_{\lambda}\Lambda_{\alpha\beta}^{\lambda}-\partial_{\beta}
\Lambda_{\alpha\lambda}^{\lambda}+\mathcal{L}_{\lambda\rho}^{\lambda}
\Lambda_{\alpha\beta}^{\rho}-\mathcal{L}_{\beta\rho}^{\lambda}\Lambda
_{\alpha\lambda}^{\rho}+\Lambda_{\rho\lambda}^{\lambda}\mathcal{L}
_{\beta\alpha}^{\rho}-\Lambda_{\rho\beta}^{\lambda}\mathcal{L}_{\lambda\alpha
}^{\rho}. \label{Q}
\end{eqnarray}
\endnumparts

The above mathematical formulation has been constructed as an attempt to generalize Einstein-Schr\"odinger theory \cite{Clerc, Crum}. In that work, the spacetime was supposed to be endowed with a metric structure. An application of the formalism to cosmology has been done by the author and his collaborator in \cite{AB,AB2}.
\newline
Despite its complication, the formalism reviewed here is solid and needs further applications. One may address a general modified theory of gravity around the new structure given here. For instance, in the framework of Bimetric theory of gravity, although we did not discuss the notion of metric here, one may define a new connection as the sum of the Christoffel symbol and the new tensor field $\Lambda_{\alpha\beta}^{\gamma}$ appeared in this formalism. In this view, the geodesic equations would be modified due to this tensor field which maybe regarded as describing the gravitational field.    
\newline
In the following section, we will study the implications of this formalism to Eddington's purely affine gravity where the possible Lagrangian densities are constructed by the Ricci-type tensors (\ref{P}) and (\ref{Q}).      
      
\section{Eddington's Theory in Immersed Spacetime}

We start our setup by defining a simple Lagrangian density which we propose to be constructed by the symmetric part of the Ricci-type tensor $\mathcal{P_{\alpha\beta}}$ given in (\ref{P}) for a symmetric connection $\mathcal{L}$. Eddington's-like gravitational action maybe taken as follows
\begin{equation}
S=\int d^{4}x \sqrt{\texttt{Det}\left[ \mathcal{P}\right]}, \label{action1}
\end{equation} 
where $\texttt{Det}\left[ \mathcal{P}\right]$ denotes the determinant of $\mathcal{P}_{\alpha\beta}$. In what follows, as in Eddington's theory, $\mathcal{P}_{\alpha\beta}$ means its symmetric part.   

As the variation will be made only with respect to the symmetric connection $\mathcal{L}$, it can be easily shown from the definition (\ref{P}) that
\begin{equation}
\delta {\mathcal{P}}_{\alpha\beta} = \nabla_{\mu}\Big( \delta\mathcal{L}^{\mu}_{\beta\alpha}\Big) -
\nabla_{\beta}\Big(\delta\mathcal{L}^{\mu}_{\mu\alpha}\Big).
\end{equation}
In this case, the variational principle applied to the action (\ref{action1}) leads to the equation of motion
\begin{equation}
\nabla_{\mu} \left[ \sqrt{{\texttt{Det}}\left[{\mathcal{P}}\right]}
\left({\mathcal{P}}^{-1}\right)^{\alpha\beta}\right]=0. \label{motion1}
\end{equation} 
As in Eddington's approach, this can be integrated by introducing an invertible and covariantly-constant tensor field $g_{\alpha\beta}$ such that
\begin{equation}
\sqrt{{\texttt{Det}}\left[{\mathcal{P}}\right]}
\left({\mathcal{P}}^{-1}\right)^{\alpha\beta}  = \lambda
\sqrt{g} g^{\alpha\beta}, \label{sol1}
\end{equation}
where $\lambda$ is a constant and $g = {\texttt{Det}}\left[g_{\alpha\beta}\right]$. The inverse $g^{\alpha\lambda}$ of the tensor $g_{\alpha\lambda}$ is such that $g^{\alpha\lambda}g_{\lambda\beta}=\delta^{\alpha}_{\beta}$.
\newline
The solution (\ref{sol1}) can be rewritten as
\begin{equation}
{\mathcal{P}}_{\alpha\beta} = \lambda g_{\alpha\beta}. \label{Eq1}
\end{equation}
As we made earlier, the compatibility condition $\nabla_{\gamma} g_{\alpha\beta} =0$ resulted from the equation of motion (\ref{motion1}) defines completely the Levi-Civita connection 
\begin{equation}
\mathcal{L}^{\mu}_{\alpha\beta}= \frac{1}{2} g^{\mu\lambda} \left( \partial_{\alpha} g_{\beta\lambda} +
\partial_{\beta} g_{\lambda\alpha} - \partial_{\lambda} g_{\alpha\beta}\right),
\end{equation}
from which the associated Ricci tensor is written as
\begin{equation}
\mathcal{R_{\alpha\beta}}=\partial
_{\lambda}\mathcal{L}_{\alpha\beta}^{\lambda}-\partial_{\alpha}\mathcal{L}%
_{\lambda\beta}^{\lambda}+\mathcal{L}_{\lambda\rho}^{\lambda}\mathcal{L}%
_{\alpha\beta}^{\rho}-\mathcal{L}_{\alpha\rho}^{\lambda}\mathcal{L}%
_{\lambda\beta}^{\rho}. \label{Ricci1} 
\end{equation}
This tensor (\ref{Ricci1}) is given by its general definition resulted from the contraction of the Riemann tensor which involves only the connection.
Now, by using (\ref{P}) and (\ref{Ricci1}), the equation (\ref{Eq1}) leads to the gravitational field equations
\begin{equation}
{\mathcal{R}}_{\alpha\beta} 
= \lambda g_{\alpha\beta}+\Lambda_{\rho (\alpha}^{\lambda}\Lambda_{\beta)\lambda}^{\rho
}-\Lambda_{\rho\lambda}^{\lambda}\Lambda_{(\beta\alpha)
}^{\rho}. \label{field1}
\end{equation}
These are Einstein's field equations including both matter and a cosmological constant $\lambda$ provided that
\begin{equation}
\Lambda_{\rho (\alpha}^{\lambda}\Lambda_{\beta)\lambda}^{\rho
}-\Lambda_{\rho\lambda}^{\lambda}\Lambda_{(\beta\alpha)
}^{\rho} = T_{\alpha\beta}-\frac{1}{2}g_{\alpha\beta} g^{\mu\nu}T_{\mu\nu},
\end{equation}
where $T_{\alpha\beta}$ is the energy momentum tensor of matter which can be written as
\begin{equation}
T_{\alpha\beta}=\left(\delta^{\nu}_{\beta}\delta^{\mu}_{\alpha}-\frac{1}{2}g_{\alpha\beta}g^{\mu\nu}\right)
\left(\Lambda_{\rho (\mu}^{\lambda}\Lambda_{\nu)\lambda}^{\rho
}-\Lambda_{\rho\lambda}^{\lambda}\Lambda_{(\nu\mu)
}^{\rho}\right). \label{matter0}
\end{equation}
As we see, matter can be reproduced from the dynamical equation (\ref{motion1}) via the new fundamental tensor $\Lambda_{\alpha\beta}^{\lambda}$ in the space. Due to the conservation of the matter tensor, one can impose a condition on this tensor by applying the divergence into equation (\ref{matter0}). One way to define an exact matter tensor from quadratic terms of $\Lambda_{\alpha\beta}^{\lambda}$ will be illustrated later in this section when we define a second action.
   
Now, let us examine the effects of the second Ricci-type tensor $\mathcal{Q}_{\alpha\beta}$ given by (\ref{Q}). Out of various possibilities, the action that contains $\mathcal{Q}_{\alpha\beta}$ can be taken as follows 
\begin{equation}
S=\int d^{4}x \sqrt{{\texttt{Det}}\left[ \mathcal{P}+\mathcal{Q}\right]}. \label{action2}
\end{equation}
From the definition of the second tensor of Ricci-type $\mathcal{Q}_{\alpha\beta}$, one can not expect a covariant-free form as in (\ref{motion1}) due to the additional linear terms in $\mathcal{L}$ which appears in $\mathcal{Q}_{\alpha\beta}$.  
In fact, variation of the action (\ref{action2}) with respect to the connection $\mathcal{L}$ gives 
\begin{eqnarray}
\fl 
\delta S=\int d^{4}x
\Bigg ( 
\nabla_{\lambda} \left[ \sqrt{{\texttt{Det}}\left[ \mathcal{S}\right]} \left(\left(\mathcal{S}^{-1}\right)^{\alpha\lambda}\delta^{\beta}_{\gamma} 
-\left(\mathcal{S}^{-1}\right)^{\alpha\beta}\delta^{\lambda}_{\gamma}\right)\right]
\nonumber\\
+\sqrt{{\texttt{Det}}\left[ \mathcal{S}\right]}\left(\mathcal{S}^{-1}\right)^{\mu\nu}
\left(\Lambda_{\mu\nu}^{\alpha}\delta^{\beta}_{\gamma} 
-\Lambda_{\mu\gamma}^{\alpha}\delta^{\beta}_{\nu}
+\Lambda_{\gamma\rho}^{\rho}\delta^{\beta}_{\nu}\delta^{\alpha}_{\mu}
-\Lambda_{\gamma\nu}^{\beta}\delta^{\alpha}_{\mu}
\right)\Bigg )\delta \mathcal{L}^{\gamma}_{\beta\alpha}, \label{varia2}
\end{eqnarray}
where we have put for simplicity $\mathcal{S}_{\alpha\beta}=\mathcal{P}_{\alpha\beta}+\mathcal{Q}_{\alpha\beta}$.
\newline
The metric structure associated to the affine structure (\ref{action2}) is obtained by introducing the symmetric tensor density
\begin{equation}
\mathcal{G}^{\mu\nu}=\lambda\sqrt{g}g^{\mu\nu}, \label{density}
\end{equation}
and by imposing the equality 
\begin{equation}
\sqrt{{\texttt{Det}}\left[ \mathcal{S}\right]} \left(\mathcal{S}^{-1}\right)^{\alpha\beta}=\mathcal{G}^{\alpha\beta}. \label{density2}
\end{equation}
When substituted into (\ref{varia2}), it leads to
\begin{eqnarray}
\delta S= \int d^{4}x \Bigg ( 
&\nabla_{\mu} \left( \mathcal{G}^{\alpha\mu}\delta^{\beta}_{\gamma}
-\mathcal{G}^{\alpha\beta}\delta^{\mu}_{\gamma}\right) \nonumber\\
&+\mathcal{G}^{\mu\nu} \left(
\Lambda_{\mu\nu}^{\alpha}\delta^{\beta}_{\gamma}
-\Lambda_{\mu\gamma}^{\alpha}\delta^{\beta}_{\nu}
+\Lambda_{\gamma\rho}^{\rho}\delta^{\beta}_{\nu}\delta^{\alpha}_{\mu}
-\Lambda_{\gamma\nu}^{\beta}\delta^{\alpha}_{\mu}
\right)\Bigg )\delta\mathcal{L}^{\gamma}_{\beta\alpha}.
\end{eqnarray}
The principle of variation $\delta S=0$ applied to the last action leads to the equation
\begin{eqnarray}
\nabla_{\mu} \left( \mathcal{G}^{\alpha\mu}\delta^{\beta}_{\gamma}
-\mathcal{G}^{\alpha\beta}\delta^{\mu}_{\gamma}\right)+\mathcal{G}^{\mu\nu} \left(
\Lambda_{\mu\nu}^{\alpha}\delta^{\beta}_{\gamma}
-\Lambda_{\mu\gamma}^{\alpha}\delta^{\beta}_{\nu}
+\Lambda_{\gamma\rho}^{\rho}\delta^{\beta}_{\nu}\delta^{\alpha}_{\mu}
-\Lambda_{\gamma\nu}^{\beta}\delta^{\alpha}_{\mu}
\right)=0. \label{densityeq}
\end{eqnarray}
When using (\ref{density}) and applying the relation 
\begin{equation}
\mathcal{G}^{\alpha\beta}\nabla_{\gamma}g_{\alpha\beta}= 2\partial_{\gamma}\sqrt{g}-2\sqrt{g}\mathcal{L}_{\lambda\gamma}^{\lambda},
\end{equation}
to (\ref{densityeq}), we can get rid of $\sqrt{g}$ and obtain in terms of the tensor $g^{\alpha\beta}$ the equation
\begin{eqnarray} 
\nabla_{\gamma} g^{\alpha\beta}
+g^{\alpha\nu}\Lambda_{\gamma\nu}^{\beta}
+g^{\mu\beta}\Lambda_{\mu\gamma}^{\alpha}- \frac{2}{3}g^{\alpha\beta} \Lambda_{\rho\gamma}^{\rho} 
+\frac{1}{3} \Lambda_{\nu} \left(g^{\alpha\nu} \delta^{\beta}_{\gamma}
- \frac{5}{3} g^{\alpha\beta} \delta_{\gamma}^{\nu}
\right) =0, \label{noncomp}
\end{eqnarray}
where we have put for simplicity $\Lambda_{\nu}=\Lambda_{\nu\rho}^{\rho}-\Lambda_{\rho\nu}^{\rho}$ which vanishes in the case of a symmetric tensor $\Lambda_{\alpha\beta}^{\gamma}$.
\newline
Unlike the compatibility equation $\nabla_{\gamma}g^{\alpha\beta}=0$, we notice the presence of the additional terms proportional to $\Lambda_{\alpha\beta}^{\gamma}$ in equation (\ref{noncomp}).
Actually, the relation (\ref{noncomp}) is an equation for the affine connection $\mathcal{L}$. In fact, for symmetric $\mathcal{L}$ which we shall call it $\Gamma$ here, the equation (\ref{noncomp}) takes the form
\begin{eqnarray}
\partial_{\gamma} g^{\alpha\beta} +g^{\alpha\lambda}\Gamma_{\gamma\lambda}^{\beta}
+g^{\lambda\beta}\Gamma_{\gamma\lambda}^{\alpha}
&=
 \frac{2}{3}g^{\alpha\beta} \Lambda_{\rho\gamma}^{\rho}
-g^{\alpha\nu}\Lambda_{\gamma\nu}^{\beta}
-g^{\mu\beta}\Lambda_{\mu\gamma}^{\alpha} 
\nonumber\\
&-\frac{1}{3} \Lambda_{\nu} \left(g^{\alpha\nu} \delta^{\beta}_{\gamma}
- \frac{5}{3} g^{\alpha\beta} \delta_{\gamma}^{\nu}
\right). 
\end{eqnarray}
A solution to this equation can be taken as \cite{Nikodem}
\begin{equation}
\Gamma_{\alpha\beta}^{\gamma}= {}^{g}\Gamma_{\alpha\beta}^{\gamma}+ K_{\alpha\beta}^{\gamma}, \label{affine2}
\end{equation}  
where ${}^{g}\Gamma_{\alpha\beta}^{\gamma}$ is the Levi-Civita connection constructed from $g_{\alpha\beta}$ and $K_{\alpha\beta}^{\gamma}$ is a tensor of rank (2,1).
\newline
One can easily check that a Ricci tensor associated to the affine connection (\ref{affine2}) can be written as
\begin{equation}
\mathcal{R_{\alpha\beta}}\left(\Gamma \right)=\mathcal{R_{\alpha\beta}}\left({}^{g}\Gamma\right)
+\nabla_{\rho}K_{\alpha\beta}^{\rho}-\nabla_{\beta}K_{\alpha\rho}^{\rho}
+K_{\alpha\beta}^{\sigma}K_{\sigma\rho}^{\rho}-K_{\alpha\rho}^{\sigma}K_{\sigma\beta}^{\rho}, \label{Ricci3} 
\end{equation}
where $\mathcal{R_{\alpha\beta}} \left({}^{g}\Gamma\right)$ is the Ricci tensor of the Livi-Civita connection ${}^{g}\Gamma$ and the covariant derivative here is with respect to ${}^{g}\Gamma$. However, this is not surprising since the Lagrangian density which we started with in (\ref{action2}) contains not only the Ricci tensor $\mathcal{R_{\alpha\beta}}\left(\mathcal{L}\right)$ but also linear terms of the connection $\mathcal{L}$.
\newline 
If we take the tensor $\Lambda_{\alpha\beta}^{\gamma}$ to be symmetric as a simple case, we can show from the relations (\ref{P}) and (\ref{Q}) that
\begin{equation}
\mathcal{P}_{\alpha\beta}+\mathcal{Q}_{\alpha\beta}=
\mathcal{R_{\alpha\beta}}\left(\mathcal{L}\right)
+\nabla_{\rho}\Lambda_{\alpha\beta}^{\rho}-\nabla_{\beta}\Lambda_{\alpha\rho}^{\rho}
+\Lambda_{\alpha\beta}^{\sigma}\Lambda_{\sigma\rho}^{\rho}
-\Lambda_{\alpha\rho}^{\sigma}\Lambda_{\sigma\beta}^{\rho}, \label{P+Q}
\end{equation}
where the covariant derivative is with respect to the connection $\mathcal{L}$.
\newline 
Comparing the two relations (\ref{P+Q}) and (\ref{Ricci3}), we directly conclude that 
\begin{equation}
K_{\alpha\beta}^{\gamma}=\Lambda_{\alpha\beta}^{\gamma},
\end{equation}
providing $\mathcal{L}={}^{g}\Gamma$.
\newline
Now, the equation (\ref{density2}) is equivalent to
\begin{equation}
\mathcal{P}_{\alpha\beta}+\mathcal{Q}_{\alpha\beta}=\lambda g_{\alpha\beta},
\end{equation}
from which we obtain the field equations when using (\ref{P+Q})
\begin{equation}
\mathcal{R_{\alpha\beta}}=\lambda g_{\alpha\beta}+\nabla_{\beta}\Lambda_{\alpha\rho}^{\rho}
-\nabla_{\rho}\Lambda_{\alpha\beta}^{\rho}+\Lambda_{\alpha\rho}^{\sigma}
\Lambda_{\sigma\beta}^{\rho}
-\Lambda_{\alpha\beta}^{\sigma}\Lambda_{\sigma\rho}^{\rho}. \label{shifted}
\end{equation}
As we see, the effect of the second tensor of Ricci-type $\mathcal{Q_{\alpha\beta}}$ is to shift the field equations (\ref{field1}) by a covariant derivative terms of the tensor $\Lambda_{\alpha\beta}^{\gamma}$. In this case, the new total term in (\ref{shifted}) that contains $\Lambda_{\alpha\beta}^{\gamma}$ forms the matter part $T_{\alpha\beta}- \frac{1}{2}g_{\alpha\beta} g^{\mu\nu}T_{\mu\nu}$. 

Although it is not trivial to construct a conserved energy momentum tensor of matter from the terms that contain $\Lambda_{\alpha\beta}^{\gamma}$ in equations (\ref{shifted}) and (\ref{field1}), one can follow a technique based on an appropriate modification of the connection $\Gamma_{\alpha\beta}^{\gamma}= {}^{g}\Gamma_{\alpha\beta}^{\gamma}+ \Lambda_{\alpha\beta}^{\gamma}$ to derive the energy momentum tensor in our case \cite{Durmus2}.
\newline
The tensor $\Lambda^{\lambda}_{\alpha\beta}$ (noted $\Delta ^{\lambda}_{\alpha\beta}$ in \cite{Durmus2}) is written in terms of a tensor $\Theta_{\alpha\beta}$ as
\begin{equation}
\label{theta}
\Lambda^{\lambda}_{\alpha\beta}= \nabla_{\alpha} \Theta_{\beta}^{\lambda}
+\nabla_{\beta} \Theta_{\alpha}^{\lambda}
-\nabla^{\lambda} \Theta_{\alpha\beta}.
\end{equation}
From this construction, the gravitational equations (\ref{shifted}) will be reduced to Einsteins field equations with matter
\begin{equation}
\mathcal{R_{\alpha\beta}}= \lambda g_{\alpha\beta} + T_{\alpha\beta}- \frac{1}{2}g_{\alpha\beta} T,
\end{equation}
when imposing
\begin{equation}
T_{\alpha\beta} = -2\left[ \mathcal{K} ^{-1}\left(\nabla \right) \right]^{\mu\nu}_{\alpha\beta} \Theta_{\mu\nu}, \label{matter}
\end{equation}
where $\left[ \mathcal{K}^{-1} \right]_{\alpha\beta\mu\nu}\left(\nabla \right)$ is the inverse propagator of a spin-2 massless field given by its expression \cite{Durmus2}
\begin{eqnarray}
\label{propagator}
\fl
\left[ \mathcal{K}^{-1} \right]_{\alpha\beta\mu\nu}\left(\nabla \right)= 
\frac{1}{8} \left( \nabla_{\mu} \nabla_{\alpha} g_{\nu\beta} + \nabla_{\mu} \nabla_{\beta} g_{\alpha\nu} \right)
+\frac{1}{8} \left( \nabla_{\nu} \nabla_{\alpha} g_{\mu\beta} + \nabla_{\nu} \nabla_{\beta} g_{\alpha\mu} \right)
\nonumber \\
-\frac{1}{8} \left( \nabla_{\alpha} \nabla_{\beta} + \nabla_{\beta} \nabla_{\alpha}  \right) g_{\mu\nu}
-\frac{1}{8} \left( \nabla_{\mu} \nabla_{\nu} + \nabla_{\nu} \nabla_{\mu}  \right) g_{\alpha\beta}
\nonumber \\
-\frac{1}{8} \opensquare \left( g_{\alpha\mu}g_{\beta\nu} + g_{\alpha\nu}g_{\mu\beta}
-2g_{\alpha\beta}g_{\mu\nu}  \right).
\end{eqnarray}

We mention here that a spin-2 massless field (or a graviton) is studied in Eddington's purely affine theory, and in Eddington-inspired Born-Infeld theory via tensor perturbation of the metric \cite{graviton1,graviton2,graviton3}. The inverse propagator (\ref{propagator}) is obtained classically without any attempt to quantization. In fact, the approach given above is just to illustrate how to construct a matter term from the general term in equation (\ref{shifted}) which include the tensor $\Lambda_{\alpha\beta}^{\gamma}$ and its covariant derivative. The resulted matter term given in (\ref{matter}) is found to be proportional to the mentioned inverse propagator. This matter term of course can not include a covariantly constant terms due to the form of this propagator, in fact any term proportional to $\lambda g_{\alpha\beta}$ which can be included in $\Theta_{\alpha\beta}$ is automatically disappeared in (\ref{matter}). For more details of how to construct such a matter term and how does the massless spin-2 propagator appear, we refer the reader to Ref.\cite{Durmus2}.

\section{Degravitating the Cosmological Constant}

A generalisation of Eddington's theory can be formulated by using a covariant Lagrangian density which can be a product of a scalar and the square root of the determinant of rank 2 tensor \cite{Schrodinger,Nikodem}. From the above study, we see that in addition to the fundamental tensor, the symmetric part of the Ricci tensor $\mathcal{R_{\alpha\beta}}$ (also the Torsion tensor in the nonsymmetric case), one can also construct covariant tensors from the new tensor $\Lambda^{\gamma}_{\alpha\beta}$.
Another possible and simple action that can be taken is
\begin{equation}
S=\int d^{4}x \sqrt{{\texttt{Det}}\left[ c_{1}\mathcal{R_{\alpha\beta}}
+c_{2}\Lambda_{\alpha\rho}^{\lambda}\Lambda_{\beta\lambda}^{\rho
}\right]},
\end{equation}
where $c_{1}$ and $c_{2}$ are constants.
\newline
Repeating the same steps given above, we obtain the gravitational field equations
\begin{equation}
c_{1}\mathcal{R}_{\alpha\beta}
+c_{2}\Lambda_{\alpha\lambda}^{\rho}\Lambda_{\beta\rho}^{\gamma}=\lambda g_{\alpha\beta},
\end{equation}
or
\begin{equation}
\mathcal{R}_{\alpha\beta}= \frac{1}{c_{1}}\lambda g_{\alpha\beta}
-\frac{c_{2}}{c_{1}}\Lambda_{\alpha\lambda}^{\rho}\Lambda_{\beta\rho}^{\lambda}. \label{CC}
\end{equation}
We can study the equations (\ref{CC}) in two different cases. First, once the term $\Lambda_{\alpha\lambda}^{\rho}\Lambda_{\beta\rho}^{\lambda}$ does not generate $\lambda g_{\alpha\beta}$ like terms, the field equations (\ref{CC}) can always be rescaled to \cite{Durmus1}  
\begin{equation}
\mathcal{R}_{\alpha\beta}= \lambda g_{\alpha\beta}
-\Lambda_{\alpha\lambda}^{\rho}\Lambda_{\beta\rho}^{\lambda}.  \label{rescaled}
\end{equation}
These are Einstein's field equations with both vacuum energy and matter defined respectively as
\numparts
\begin{eqnarray}
T^{vac}_{\alpha\beta}=-M^{2}_{Pl}\lambda g_{\alpha\beta} \label{vac} \\
-M^{2}_{Pl}\Lambda_{\alpha\lambda}^{\rho}\Lambda_{\beta\rho}^{\lambda}= T^{matt}_{\alpha\beta}- \frac{1}{2}g_{\alpha\beta} g^{\mu\nu}T^{matt}_{\mu\nu}, \label{matt}
\end{eqnarray}
\endnumparts
where $M^{2}_{Pl}=\left( 8\pi G_{N}\right)^{-1}$ is the Planck mass, and the conservation of the energy momentum tensor of matter imposes a condition on the tensor $\Lambda^{\gamma}_{\alpha\beta}$ as
\begin{equation}
2\nabla^{\alpha} \left( \Lambda_{\alpha\lambda}^{\rho} \Lambda_{\beta\rho}^{\lambda} \right)
- g^{\mu\nu} 
\nabla_{\beta} \left( \Lambda_{\mu\lambda}^{\rho} \Lambda_{\nu\rho}^{\lambda} \right)=0.
\end{equation}
This first case given by the rescaling (\ref{rescaled}) is classical, it is supposed that matter does not undergo phase transitions that releases vacuum energy which shall contribute again to the cosmological constant $\lambda$. We note here that there are other kinds of phase transitions which can be seen in the framework of Eddington-inspired Born Infeld theory, where in addition to a static Einstein's phase, the universe undergoes a regular bounce \cite{Banados}. However, phase transitions that we mentioned here are completely different, and they are in a sense of particle theory. 
\newline
However, the first case is not realistic, phase transitions should have been occurring and then matter might develop quantum corrections of the vacuum that should be always added to the cosmological constant. This translates the second case in which the terms $\Lambda_{\alpha\lambda}^{\rho}\Lambda_{\beta\rho}^{\lambda}$ and $\lambda$ are not independent, then the rescaling made to obtain equation (\ref{rescaled}) fails.   
\newline
Nevertheless, equation (\ref{CC}) can be written using (\ref{vac}) and (\ref{matt}) as  
\begin{equation}
\mathcal{R_{\alpha\beta}}=\left(\frac{M_{Pl}^{-2}}{c_{1}} \right) \mathcal{E} g_{\alpha\beta}
+\left(\frac{c_{2}}{c_{1}}\right) M_{Pl}^{-2}\left(T^{matt}_{\alpha\beta}- \frac{1}{2}g_{\alpha\beta} g^{\mu\nu}T^{matt}_{\mu\nu}\right), \label{CC2}
\end{equation}
where $\mathcal{E}=M^{2}_{Pl} \lambda $ is the vacuum energy density.
\newline
The constants $c_{1}$ and $c_{2}$ are arbitrary and we are free to make some constraint due to physical reasons if any. In General Relativity, matter and radiation gravitate with the ordinary Newton's constant $8 \pi G_{N}=M_{Pl}^{-2}$, then any modified theory of gravity should fit this constraint in order not to affect the standard astrophysics and cosmology \cite{Durmus1, Nima}. To ensure this, we impose the condition $c_{2}=c_{1}$ on equation (\ref{CC2}) and then we get
\begin{equation}
\mathcal{R_{\alpha\beta}}=\left(\frac{M_{Pl}^{-2}}{c_{1}} \right) \mathcal{E} g_{\alpha\beta}
+ M_{Pl}^{-2}\left(T^{matt}_{\alpha\beta}- \frac{1}{2}g_{\alpha\beta} g^{\mu\nu}T^{matt}_{\mu\nu}\right). \label{CC3}
\end{equation}
Like matter and radiation, vacuum energy appears to gravitate with the Newton's constant $G_{N}$ (or $M_{Pl}^{-2}$) in Einstein's General Relativity. One of the sources of this vacuum energy are the zero-point energies of quantum fields. These energies appear to be of the order $\Lambda_{UV}^{4}$, where $\Lambda_{UV}$ is the Ultra-Violet Cutoff which is the scale up to which the field theory is valid. This theoretical value seem to cause a perplexed problem when is viewed in the framework of General Relativity. In fact, spacetime is extremely sensitive to this vacuum. If we trust quantum field theory up to Planck scale, $\Lambda_{UV}=M_{Pl}$, the mentioned vacuum energy causes empty space to possess a curvature of the order $\mathcal{R} \sim M_{Pl}^{2}$ which appears to be about 120 orders of magnitude larger than the observed value $\mathcal{R} \sim \frac{m_{\nu}^{4}}{M_{Pl}^{2}}$, where $m_{\nu}$ is the Neutrino mass \cite{Planck}. This is called the cosmological constant problem, the most intriguing problem in General Relativity and Particle Physics \cite{Weinberg,Sahni}.
\newline
As we see from equation (\ref{CC3}), vacuum energy gravitates with the new strength $\frac{M_{Pl}^{-2}}{c_{1}}$, and up to now the constant $c_{1}$ is arbitrary and it should be fixed by some physical conditions if any. Unfortunately, our model does not offer any theoretical derivation to fix this constant. Nevertheless, one may argue that the very small curvature maybe recovered for a very large constant $c_{1}$. In addition to the planck mass scale, one can introduce a new cosmological strength $M_{Co}$ as another scale such that the constant $c_{1}$ is defined as the ratio of these two hierarchically mass scales as in the philosophy of Dirac's larger number hypothesis \cite{Dirac}, then we propose the following form
\begin{equation}
\label{ratio}
c_{1}= \left(\frac{M_{Co}}{M_{Pl}}\right)^{2}.
\end{equation}
The obtained field equations (\ref{CC3}) become
\begin{equation}
\mathcal{R_{\alpha\beta}}= M_{Co}^{-2} \mathcal{E} g_{\alpha\beta}
+ M_{Pl}^{-2}\left(T^{matt}_{\alpha\beta}- \frac{1}{2}g_{\alpha\beta}T^{matt}\right), \label{CC4}
\end{equation}
where $T^{matt}=g^{\mu\nu}T^{matt}_{\mu\nu}$.
As a result, the vacuum energy density $\mathcal{E}$ which receives its value from contributions of the zero point energies of quantum fields as well as from phase transitions, appears here to gravitate with the new gravitational constant $M^{-2}_{Co}$ and then it curves the space by the amount
\begin{equation}
\label{cur}
\mathcal{R}=4M^{-2}_{Co} \mathcal{E},
\end{equation}
rather than $4M^{-2}_{Pl} \mathcal{E}$ as in General Relativity.
The new mass scale $M_{Co}$, which is introduced here to define the ratio (\ref{ratio}), can be fixed by using the observational bounds discussed earlier which provide that
\begin{equation}
M_{Co}^{2} \simeq \lambda \left(\frac{M_{Pl}}{m_{\nu}}\right)^{4}.
\label{MCo}
\end{equation}
Thus, the observed curvature $\mathcal{R} \sim \frac{m_{\nu}^{4}}{M_{Pl}^{2}}$, which can be obtained here from (\ref{cur}) and (\ref{MCo}), is now reproduced by the hierarchy between the two scales $M_{Co}$ and $M_{Pl}$ rather than by fine-tuning.
\newline
To define the new mass scale $M_{Co}$, we followed almost the same technique used in a recent model named ``Riemann-improved Eddington theory'' \cite{Durmus1}.  But we should note here that although it leads to the same conclusion about degravitating the cosmological constant as in \cite{Durmus1}, the action we proposed here is completely different. The common point is how to propose the new mass scale (\ref{MCo}) in the philosophy of Dirac's larger number hypothesis.     
\newline
Although it enables degravitation of the vacuum energy to its observed value without fine-tuning, the model (\ref{CC3}) does not solve the cosmological constant problem. This is because the scale $M_{Co}$ is given by an empirical relation (\ref{MCo}) and one has to derive it from dynamics. 
   
\section{Summary}
The mean goal of this paper was to incorporate matter in a purely affine gravity by reformulating Eddington's theory in an immersed spacetime.
\newline
The spacetime in this model was supposed to be plunged into an octo-dimensional space which has a pseudo-complex structure. Due to this immersion, the space is endowed with a new tensor of rank (2,1) in addition to the affine connection. We have proposed different possible constructions of the gravitational action and studied in details the resulting field equations. We found that the gravitational field equations with matter can be recovered from the dynamical equations due to the presence of the new tensor of rank (2,1) in the expressions of the Ricci-type tensors.

As an attempt to degravitating the vacuum energy from this model, we have proposed a Lagrangian density as a combination of the standard Ricci tensor of the affine connection as well as a term proportional to the new rank (2,1) tensor. As a result, we found that in contrast to General Relativity where both vacuum and matter gravitate with Newton's constant, the vacuum energy gravitates here with a different cosmological strength $M_{Co}$. In this setup, without fine-tuning, the observed curvature is reproduced by the hierarchy between the Planck mass scale and $M_{Co}$.   

We conclude by noticing that the work done in this paper may not be restricted to a symmetric affine connection as in the original Eddington's theory, then one can generalize the model by extending it to involve Torsion tensor and study its effects on the field equations.      
\section*{Acknowledgements}
I am very thankful to Durmu{\c s} Ali Demir for many valuable discussions and suggestions during the preparation of this work.

\end{document}